\documentclass[aps,twocolumn,pra,showpacs,floatfix]{revtex4-1}

\usepackage{amssymb}
\usepackage{amsmath}
\usepackage{graphicx}

\begin{document}

\title{\bf QED radiative corrections and many-body effects in atoms:
vacuum polarization and binding energy shifts in alkali metals}
\author{J. S. M. Ginges}
\affiliation{School of Physics, University of New South Wales,
Sydney NSW 2052, Australia}
\author{J. C. Berengut}
\affiliation{School of Physics, University of New South Wales,
Sydney NSW 2052, Australia}

\date{\today}

\begin{abstract}

We calculate vacuum polarization corrections to the binding energies
in neutral alkali atoms Na through to the superheavy element E119. 
We employ the relativistic Hartree-Fock method to
demonstrate the importance of relaxation of the electronic core  
and the correlation potential method to study the effects of second and
higher orders of perturbation theory.
These many-body effects are sizeable for all orbitals, though particularly
important for orbitals with angular momentum quantum number $l>0$. 
The orders of magnitude enhancement for $d$ waves produces
shifts that, for Rb and the heavier elements, are larger than those
for $p$ waves and only an order of
magnitude smaller than the $s$-wave shifts. 
The many-body enhancement mechanisms 
that operate for vacuum polarization 
apply also to the larger self-energy corrections. 

\end{abstract}


\maketitle

\section{Introduction}
 
Within the last decade 
the need has arisen for a method that 
incorporates 
quantum electrodynamics (QED) radiative corrections into the many-body problem 
for heavy atoms and ions. For example, account of QED radiative corrections, that is, 
the vacuum polarization and self-energy
corrections, were critical in the interpretation of the measurement \cite{wieman} of
the parity violating amplitude in cesium, which restored an apparent
deviation from the standard model of particle physics (see, e.g., 
\cite{Sush_ueh,JBS_ueh,KF2002,MST2002,SPVC2003,SPTY2005,rad_pot,review}).
There are other areas where full account of QED and many-body effects are
required, including in the transition frequencies of many-electron highly-charged
ions (see, e.g., \cite{blundell,CCJS}) and in the spectra and chemical properties of superheavy
elements \cite{Schwerdt_review}. 
Ongoing improvements in experimental precision and developments 
in many-body methods across a range of precision atomic applications 
necessitate the accurate treatment of combined many-body and QED
effects in the generic atomic problem.

The one-loop radiative corrections are made up of the (local) vacuum
polarization and the (non-local) self-energy terms. The vacuum
polarization is dominated by the (lowest order in $Z\alpha$) Uehling
potential, and incorporation of this potential into the many-body
problem is relatively 
straightforward.

The rigorous, highly-accurate {\it ab initio} procedures for  
calculating QED self-energy corrections to properties of hydrogen-like and
few-electron atoms and ions cannot be applied with the same success to more
complex many-body systems (see, for example, the review \cite{mohr_review}). 
There are exact QED calculations performed in frozen atomic
potentials (see, e.g., \cite{labzowsky,sapcheng}), though here the many-body 
aspect of the problem 
may not be readily determined. On the other hand, 
the QED part of the problem may be simplified and the full 
many-body treatment employed. Several such ``radiative potentials'' have 
been advocated for use in many-body calculations. These include the 
local potentials \cite{pyykko,rad_pot} and the more sophisticated
non-local potential \cite{shabaev}, all fitted to self-energy shifts in hydrogenlike
ions.
 
There are a number of recent works where QED radiative 
corrections have been included into the many-body problem 
using one or other of these 
radiative potentials (e.g.,
\cite{thierfelder,roberts,radium,nature}). 
However, there has as yet been no in-depth study of the interplay 
between radiative corrections and many-body effects or of the breakdown 
of different terms. Therefore, it is important to demonstrate 
the significance of different many-body effects so that they may be 
considered appropriately in high-precision calculations.

Derevianko {\it et al.} \cite{derev_relax} studied the effects of core
relaxation on 
valence $s$ and $p$ level shifts of neutral Cs due to the
Uehling potential in the relativistic 
Hartree-Fock approximation. They demonstrated how important 
these relaxation effects are, particularly for orbitals with orbital 
quantum number $l>0$.
In this paper we extend the work of Derevianko {\it et
  al.} \cite{derev_relax} to include the effects on valence 
$d$ levels. 
Inclusion of relaxation corrections leads to several orders 
of magnitude enhancement of the Uehling correction to the $d$-level 
binding energies. 

We study also the effect of second- and higher-order many-body
perturbation theory on the 
Uehling correction to valence binding energies through inclusion of
the second-order and all-orders correlation potentials. 
We have found significant corrections to Uehling shifts for all
orbitals. 

The underlying
mechanisms leading to the large many-body effects are studied and
calculations are performed for the series of alkali-metal atoms 
Na through to the superheavy element E119.
This work serves as a preliminary study of the more general 
QED-many-body interplay, where the full radiative corrections
(including the larger self-energy) are considered. 
Since the Uehling potential may be determined to very high precision 
and is a local operator which may readily be included into atomic 
many-body codes, the study of the many-body effects is not 
hindered by any complexity arising from the QED terms themselves.
While the form of
the self-energy terms is more sophisticated than Uehling, and the
radiative shift is generally larger and of opposite sign, the short range of the
interaction means that the same many-body enhancement
mechanisms apply.

The structure of the paper is the following. In Section
\ref{sec:finite} we present the finite-size form of the Uehling
potential used in this work. In Section \ref{sec:first}, 
the first-order valence shifts are calculated for frozen 
atomic cores corresponding to core-Hartree, Kohn-Sham, 
and Hartree-Fock approximations. The many-body effects 
of core relaxation and valence-core correlations are studied 
in Sections \ref{sec:relax} and \ref{sec:correlations}, respectively. 
In these sections, details of the many-body enhancement mechanisms 
are presented and calculations performed for Na through to E119, 
with more detailed results given for Cs.

\section{Finite-size Uehling potential}
\label{sec:finite}

The Uehling potential \cite{uehling} gives the lowest-order in $Z \alpha$ vacuum
polarization correction to the Coulomb potential $V _{\rm nuc}({\bf r})$; $Z$ is
the nuclear charge and $\alpha$ is the fine-structure constant. For the
point-nucleus case, where $V_{\rm nuc}^{\rm point}(r)=Z/r$, the 
Uehling potential may be expressed as
\begin{equation}
V_{\rm Ueh}^{\rm point} (r) = \frac{2}{3}\Big(\frac{\alpha}{\pi}\Big)\Big(\frac{Z}{r}\Big) \int_1^{\infty}dt
\sqrt{t^2-1}\Big(\frac{1}{t^2}+\frac{1}{2t^4}\Big)e^{-2tr/\alpha} \ .
\end{equation}
Here and throughout we 
use atomic units ($\hbar=e=m=1$, $c=1/\alpha$). 

To obtain the finite-nucleus expression for the Uehling potential, the 
expression for the point-like nucleus is folded with the nuclear
density $\rho_{\rm nuc}$,
\begin{equation}
V_{\rm Ueh}^{\rm fin}({\bf r})=\frac{1}{Z}\int d^3r' V_{\rm Ueh}^{\rm point}(|{\bf
    r}-{\bf r}'|) \rho_{\rm nuc} ({\bf r}') \ ,
\end{equation}
where we use the normalization $\int \rho_{\rm nuc} ({\bf r})d^3r=Z$.
Taking the nuclear density to be spherically symmetric, and after angular
integrations, the well-known finite-nucleus 
expression for the Uehling potential is obtained
\begin{widetext}
\begin{equation}
\label{eq:double}
V_{\rm Ueh}^{\rm fin}(r)=\frac{2}{3}\frac{\alpha^2}{r}\int_0^{\infty}dr'\int_1^{\infty}dt\sqrt{t^2-1}\Big(
\frac{1}{t^3}+\frac{1}{2t^5}\Big) \rho_{\rm nuc} (r') r' \Big(
e^{-2t|r-r'|/\alpha}-e^{-2t(r+r')/\alpha}
\Big) \ .
\end{equation}
\end{widetext}
There are publicly available procedures for performing
this integral. For example, the code of Hnizdo \cite{hnizdo} or the expansion of
Fullerton and Rinker \cite{FR}. We will instead consider the case of the
step-function density (homogeneous charge distribution) and reduce the 
double integration, Eq.~(\ref{eq:double}), to a single
integration. The Uehling potential then
reduces to the following form, divided into parts valid inside and
outside the nucleus, 
\begin{widetext}
\begin{equation}
\label{eq:step}
V_{\rm Ueh}^{\rm step}(r) =
\Bigg\{
\begin{array}{ll}
\frac{\alpha}{\pi} \frac{Z}{r}
\int_1^{\infty}dt \sqrt{t^2-1} \Big( \frac{1}{t^2} +\frac{1}{2t^4}
\Big) \Big(\frac{2}{\varkappa^3}\Big)\Big[ \frac{r}{r_n}\varkappa -
e^{-\varkappa}(1+\varkappa) \sinh(2tr/\alpha)\Big] \ , & \qquad r\leq 
r_n \\
\frac{\alpha}{\pi}\frac{Z}{r}\int_1^{\infty}dt
\sqrt{t^2-1}\Big(\frac{1}{t^2}+\frac{1}{2t^4}\Big) \Big(\frac{2}{\varkappa^3}\Big) e^{-2tr/\alpha}
\Big[ \varkappa \cosh (\varkappa)-\sinh(\varkappa)\Big] \ , & \qquad
r> r_n \ \ ,
\end{array}
\end{equation}
\end{widetext}
where $r_n$ is the nuclear radius and $\varkappa=2 t r_n/\alpha$. 
The first term inside the square brackets of the upper line of Eq.~(\ref{eq:step}) may be
simplified by integrating over $t$ analytically, yielding 
$\frac{\alpha}{\pi}\frac{Z}{r}\big(\frac{\alpha}{2r_n}\big)^3\Big(\frac{8r}{5\alpha}\Big)$,
where all factors have been included. Eq.~(\ref{eq:step})
agrees with that given in Ref. \cite{huang} in a different form.

Throughout the paper we take $V_{\rm Ueh}(r)=V_{\rm Ueh}^{\rm
  step}(r)$ and $r_n=\sqrt{5/3}\, r_{\rm rms}$, where the nuclear
root-mean-square radii $r_{\rm rms}$ are taken from Ref.~\cite{rms}.
We have found that there is agreement to all digits presented for binding-energy shifts
calculated using the step-function
density expression for the Uehling potential, Eq. (\ref{eq:step}), compared to the full
double integral Eq.~(\ref{eq:double}) with a two-parameter Fermi
distribution used for the nuclear density $\rho_{\rm nuc}(r)$,
provided the same root-mean-square nuclear radius is used.

In this work the integration of Eq. (\ref{eq:step}) is performed 
using the GNU Scientific Library adaptive integration routine QAGI \cite{gsl}.

\section{First-order shifts}
\label{sec:first}

The first-order Uehling correction to the binding energies of the
valence electron is given by
\begin{equation}
\delta \epsilon_i^{(1)}=-\langle \varphi_i |V_{\rm Ueh}|\varphi_i\rangle \ ,
\end{equation}
where $\varphi_i$ is the valence electron wave function of state $i$.

The zeroth-order valence energies $\epsilon_i$ and wave functions
$\varphi_i$ are found by solving the relativistic equations
\begin{equation}
\label{eq:rel}
( c\boldsymbol{\alpha}\cdot {\bf p} +(\beta -1)c^2 -V_{\rm nuc}-V_{\rm
  el}) \varphi_i= \epsilon_i \varphi_i \ ,
\end{equation}
where $V_{\rm nuc}$ and $V_{\rm el}$ are the nuclear and electronic
potentials, $\boldsymbol{\alpha}$ and $\beta$ are Dirac matrices, and 
${\bf p}$ is the momentum operator. We take the
nuclear density to correspond to the two-parameter Fermi
distribution, with the nuclear thickness (90\% to 10\% fall-off) taken to be
$2.3\, {\rm fm}$ for all atoms and the half-density radii found from
the root-mean-square radii of Ref.~\cite{rms}. For E119, we take $r_{\rm rms}=6.5\, {\rm fm}$, within the range predicted
from Hartree-Fock-BCS theory \cite{BCS}.

\begin{table*}
\caption{First-order Uehling corrections $\delta \epsilon^{(1)}$ to
  $s$-wave binding energies in core-Hartree, Kohn-Sham, and
  Hartree-Fock approximations. 
Zeroth-order core-Hartree and Kohn-Sham energies, $\epsilon_{CH}$
and $\epsilon_{KS}$, are also
presented. Vacuum polarization results of Ref. \cite{sapcheng}
are shown.  
The numbers in square brackets $[\,]$ denote powers of
10. Units: a.u.}
\label{tab:CHKS}
\begin{ruledtabular}
\begin{tabular}{llccccccc}
Atom &State &$\epsilon_{\rm CH}$&
\multicolumn{2}{c}{$\delta\epsilon_{\rm CH}^{(1)}$}
&$\epsilon_{\rm KS}$&\multicolumn{2}{c}{$\delta\epsilon_{\rm
    KS}^{(1)}$}& $\delta\epsilon_{\rm HF}^{(1)}$\\
&& &This work & Ref.~\cite{sapcheng}& &This work & Ref.~\cite{sapcheng} &
This work\\
\hline
Na &$3s_{1/2}$&-0.173341& -7.056[-7] & -6.888[-7] &-0.178764& -6.678[-7] & -6.573[-7] & -5.559[-7] \\
K& $4s_{1/2}$ &-0.139522&-1.497[-6]& -1.472[-6]&-0.144032&-1.431[-6] & -1.416[-6] & -1.224[-6]\\
Rb & $5s_{1/2}$ &-0.131786&-5.274[-6] &-5.224[-6] &-0.135854&-5.202[-6]&-5.167[-6] & -4.648[-6]\\
Cs & $6s_{1/2}$&-0.120057& -1.131[-5]& -1.123[-5] &-0.124015&-1.133[-5] & -1.128[-5] &  -1.054[-5]\\
Fr& $7s_{1/2}$ &-0.122284 & -4.556[-5]&-4.533[-5]&-0.125432&-4.642[-5] & -4.628[-5] & -4.997[-5]\\
\end{tabular}
\end{ruledtabular}
\end{table*}

In this work we consider three different electronic potentials $V_{\rm
  el}$. One is the relativistic Hartree-Fock potential, which we use
as our starting point for many-body perturbation theory. We also consider the simpler 
core-Hartree and Kohn-Sham potentials as points of reference. 
Labzowsky {\it et al.} \cite{labzowsky} and Sapirstein and Cheng \cite{sapcheng} have calculated the $s$-wave
vacuum polarization (and self-energy) corrections for the alkali metals using several 
atomic potentials. The core-Hartree and Kohn-Sham potentials were those
favored in Ref.~\cite{sapcheng} from the five atomic potentials, based
on local density theory, considered in that work.

In the Hartree-Fock approximation, Eq.~(\ref{eq:rel}) is first
solved self-consistently for all orbitals of the core, and 
$V_{\rm el}$ represents the direct and exchange Hartree-Fock
potentials of the core electrons. The core is then frozen, and
the energies and wave functions for the valence electron are found in
this potential. (Explicit expressions for the relativistic
Hartree-Fock potentials may be found in, e.g., Ref.~\cite{johnson_book}.)

In the core-Hartree approximation, the procedure is similar to that for
Hartree-Fock, though the exchange potential is
altogether excluded. In the Kohn-Sham approximation, a local form for
the exchange potential is used,
$V_{\rm exch}^{\rm KS}(r)=\frac{2}{3}( \frac{81}{32\pi^2}
\rho_{\rm el}(r))^{1/3}$, where $\rho_{\rm el}(r)$ is the total (core
and valence) electronic density, with each
electron contributing $\rho_i(r)=(f_i(r)^2+\alpha^2 g_i(r)^2)/r^2$.  
(The upper and lower radial components of each orbital, $f_i$ and $g_i$,
respectively, are normalized according to $(f_i^2+\alpha^2g_i^2)dr=1$.) 
The Latter correction \cite{latter} is
enforced to give the orbitals the correct asymptotic
behaviour: the total potential 
$V_{\rm tot}=V_{\rm nuc}+V_{\rm
el}$ is set to $1/r$ when $V_{\rm tot}$ falls below $1/r$.
In the Kohn-Sham approximation, Eq.~(\ref{eq:rel}) is solved
self-consistently for the core and valence electrons together. 

In Table~\ref{tab:CHKS} we present our results for the first-order
Uehling corrections to the $s$-wave valence binding energies for the alkali atoms Na
through to Fr performed in the core-Hartree and Kohn-Sham approximations; 
we also present our first-order Hartree-Fock results for comparison 
(a more comprehensive list of Hartree-Fock results is presented later in the paper in
Table~\ref{tab:all-order}). 
Our core-Hartree and Kohn-Sham results are presented alongside the 
vacuum polarization calculations of Sapirstein and Cheng
\cite{sapcheng} performed in the same potentials.
As well as taking into account all-orders in $Z\alpha$ in the vacuum
polarization, the authors of Ref.~\cite{sapcheng} include
electronic screening through the use of an effective charge. Both
effects act to reduce the size of the Uehling correction.
This may explain the deviation between the results.

The core-Hartree and Kohn-Sham results are in fairly good agreement
with each other, though 
reference to the results performed in other atomic potentials
\cite{sapcheng,labzowsky} shows that there is a sizeable spread in the
values, on the order of 10\% the size of the shifts. There is also a
significant difference between the local atomic potential results and
those performed in Hartree-Fock, see Table~\ref{tab:CHKS}. It
illustrates how sensitive the vacuum polarization corrections are to
the details of atomic structure.

\section{Core relaxation}
\label{sec:relax}

The Uehling potential affects the binding energies of the valence
electron not only through the direct first-order shift $\delta
\epsilon ^{(1)}$, given above, 
but also indirectly through Uehling corrections to the core
electrons. All-orders account of the Uehling correction in the core may 
be easily determined by adding the Uehling potential to the nuclear 
potential, $V_{\rm nuc}+V_{\rm Ueh}$, in the relativistic Hartree-Fock
equations. Self-consistent solution for the electrons of the core then
leads to a new Hartree-Fock potential 
$V^{\rm Ueh}_{\rm HF}$ which contains a correction due to Uehling  
$\delta V_{\rm HF}^{\rm Ueh}=V_{\rm HF}^{\rm Ueh}-V_{\rm HF}$, where 
$V_{\rm HF}^{\rm Ueh}$ and $V_{\rm HF}$ are, respectively, the self-consistent relativistic Hartree-Fock
potentials with and without the Uehling potential included. 
This correction, $\delta V_{\rm HF}^{\rm Ueh}$, is
referred to as a {\it relaxation} correction.
The correction to the energy of the valence electron may then be
expressed as 
\begin{equation}
\label{eq:relax_me}
\delta \epsilon_i=-\langle \varphi_i |V_{\rm Ueh}+\delta
V_{\rm HF}^{\rm Ueh}|\varphi_i\rangle =\delta
\epsilon_i^{(1)}+ \delta \epsilon_i^{\rm relax} \ .
\end{equation}

While the source of the relaxation correction is made more transparent
from inspection of Eq.~(\ref{eq:relax_me}), in calculations we instead
find the energies $\epsilon_i ^\prime $ from the solution of
the equation
\begin{equation}
\label{eq:relax1}
(c\boldsymbol{\alpha}\cdot {\bf p} +(\beta -1)c^2 -V_{\rm nuc}-V_{\rm
  Ueh} -V_{\rm HF}^{\rm Ueh})\varphi_i^\prime= \epsilon_i^\prime \varphi_i^\prime \ .
\end{equation}
The correction is then
\begin{equation}
\label{eq:relax2}
\delta \epsilon_i= \epsilon^\prime_i - \epsilon_i \ .
\end{equation}
It should be noted that the energies from the matrix element
Eq.~(\ref{eq:relax_me}) and from
Eqs.~(\ref{eq:relax1}), (\ref{eq:relax2}) are not equivalent; energies
from the solution of the latter equations include higher-order
corrections in Uehling to the valence orbitals, while the former
equation does not. 

In this and the following section we use Cs as our main atom of 
interest in consideration of the many-body effects core polarization
and valence-core correlations, see Tables~\ref{tab:deltaV} and
\ref{tab:sigma}.  Our final values for Cs, as well as for the other
alkali atoms Na through to E119, are presented in Table~\ref{tab:all-order}.

\begin{figure}[tbp]
\begin{center}
\includegraphics[width=\columnwidth]{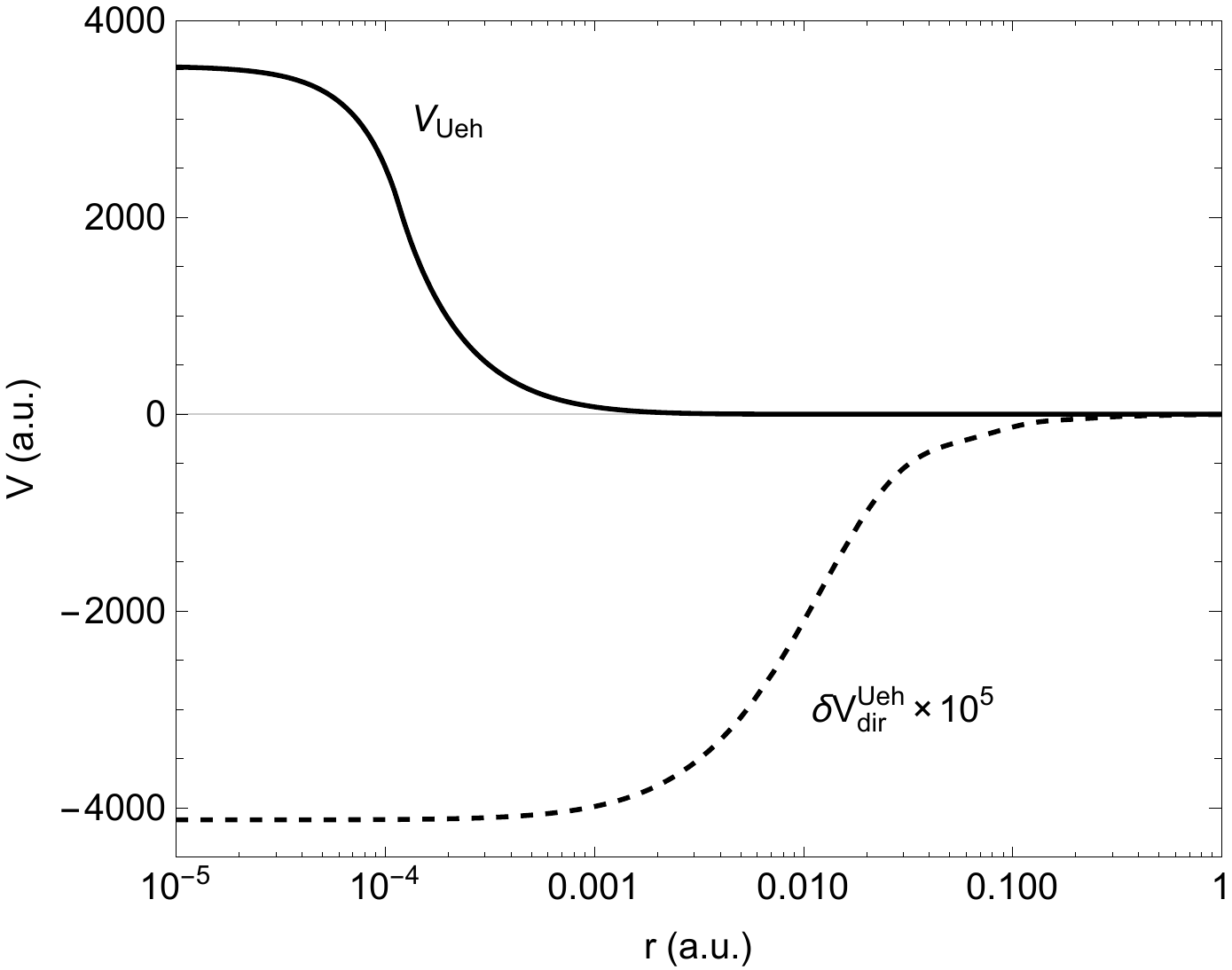}
\caption{The Uehling potential $V_{\rm Ueh}$ and the Uehling
  correction to the direct 
relativistic Hartree-Fock potential $\delta V_{\rm dir}^{\rm Ueh}$ magnified by $10^5$;
  the potentials are for Cs.}
\label{fig:potentials}
\end{center}
\end{figure}

In Fig.~\ref{fig:potentials} we present both the Uehling potential and the
Uehling correction to the Hartree-Fock direct potential for Cs. The long range of the
relaxation potential makes it possible for all orbitals to receive a
relatively sizeable shift from this mechanism. 
For orbitals with $l>0$, the relaxation contribution may be orders of magnitude
larger than the first-order Uehling contribution, 
since these orbitals do not penetrate the region close to the nucleus
where the Uehling potential acts, making the first-order contribution 
very small.

\begin{table*}
\caption{Uehling shifts for Cs. First-order valence shifts are shown
  in column two. Individual contributions of core $s$ orbitals to relaxation shifts are given in columns 3 to 7, contributions
  from all core $s$, all core $p_{1/2}$, and all core $p_{3/2}$ are presented in columns 8 to
  10. 
In the final two columns the total
 direct and exchange parts of the core relaxation shifts are given. 
Units: ${\rm  a.u.}$}
\label{tab:deltaV}
\begin{ruledtabular}
\begin{tabular}{lccccccccccc}
State &$\langle \varphi |-V_{\rm Ueh}|\varphi\rangle$ &
\multicolumn{10}{c}{Contributions to the relaxation shift, $\delta \epsilon^{\rm relax}\times 10^{6}$}\\ 
&$\times 10^{6}$&$1s$&$2s$&$3s$&$4s$&$5s$ & core $s$ & core $p_{1/2}$ & core
$p_{3/2}$& $\langle -\delta V_{\rm dir}^{\rm Ueh}\rangle$ & $\langle -\delta V_{\rm exch}^{\rm Ueh}\rangle$ \\
\hline
$6s_{1/2}$& -10.538&-0.033 &-0.139&-0.164&-0.247&-0.619&-1.211&0.116&0.035&1.573&-2.623\\
$6p_{1/2}$ &-0.194&0.040&0.031&0.027&0.048&0.284&0.430&-0.019&0.012&0.608&-0.185\\
$6p_{3/2}$ &-0.022&0.024&0.026&0.020&0.043&0.315&0.428&0.042&0.004&0.570&-0.096\\
$5d_{3/2}$&0.000&-0.104&-0.022&0.046&0.081&0.951&0.953&0.153&0.022& 0.737&0.390\\
$5d_{5/2}$&0.000&-0.098&-0.025&0.046&0.083&0.901&0.907&0.068&0.031&0.709&0.298\\
\end{tabular}
\end{ruledtabular}
\end{table*}

The first-order Uehling shifts and the breakdown of the core
relaxation contributions for the lowest five valence levels of Cs 
are presented in Table~\ref{tab:deltaV}. In the second column the 
first-order results $\delta \epsilon^{(1)}$ are given. 
In the final two columns, the shifts due to core  
relaxation are presented, divided into the direct and exchange
contributions. 

The first-order shift $\delta \epsilon^{(1)}$ is always negative,
leading to increased binding, and shifts for the
$p$ and $d$ waves are seen to be orders of magnitude smaller than the $s$-wave
shift. 
The direct part of the relaxation potential is of opposite sign to
Uehling (see also Fig.~\ref{fig:potentials}), reducing the binding. The exchange
part of the relaxation potential, however, may lead to shifts that are
positive or negative. It is interesting that for $6s$, the exchange
relaxation contribution is larger than that coming from the direct
part. For the other orbitals, we see that the direct part of the
relaxation correction dominates, though the exchange part is also 
very important, particularly for $5d$.
Consideration of core relaxation makes the Uehling
shift for $6s$ larger, while it leads to a correction of opposite sign
for the other orbitals. 

While the magnitude of the relaxation
correction for $6s$ is about 10\% that of the first-order shift, 
for the other orbitals the situation is very different. For $6p_{1/2}$,
the relaxation correction is about twice the size of the first-order shift,
which then gives a total shift that is roughly the same size as that 
of the first-order, though of opposite sign. For $6p_{3/2}$, the size
of the core relaxation correction is of similar magnitude to that for
$6p_{1/2}$, though the smaller first-order correction leads to a much 
larger (about 20 times) relative correction. For $5d_{3/2}$ and
$5d_{5/2}$, the relative size of the relaxation corrections is
enormous; the absolute corrections are about the same size as that for $6s$, 
which makes the Uehling shifts for $5d_{3/2}$ and $5d_{5/2}$ as much
as about 10\% the size of the $6s$ shift, and of opposite sign. 

While the relative sizes of the relaxation corrections for $4f$ and $5g$
orbitals are orders of magnitude larger still, the absolute sizes
of the shifts are less than $10^{-3}$ the size of the $5d$
shifts, so we will not consider the shifts for $f$ and $g$ levels further.

We have looked at the contributions to the direct and exchange
relaxation potentials, $\langle \varphi |-\delta V_{\rm dir}|\varphi
\rangle$ and $\langle \varphi |-\delta V_{\rm exch}|\varphi
\rangle$, for Cs coming from the intervals $r\le a_B$ and $r>a_B$. 
As we might expect, from examination of 
Fig.~\ref{fig:potentials}, most of the
direct relaxation shift comes from the interval $r\le a_B$.
While for $6s$ most of the exchange
relaxation shift comes from $r\le a_B$, for $6p$ and $5d$ the largest part of the exchange
relaxation shift comes from $r> a_B$.

It is interesting to see how Uehling corrections to individual
orbitals of the core propagate in the self-consistent solution of the 
corrected Hartree-Fock potential and appear finally in the shift in the valence
binding energies. In Table~\ref{tab:deltaV} in the columns $3-10$, we 
consider such shifts. In the columns under $1s$ through to $5s$, we 
present the shifts to the valence levels that result by switching on 
the Uehling correction to these core $s$ orbitals individually.
(Note that in this case the potentials for all $s$ orbitals are not
the same, resulting in orbitals that are slightly non-orthogonal.)

In columns $8-10$ of Table~\ref{tab:deltaV} we show the contributions 
to the valence shifts that result by including the Uehling potential for 
all core $s$ orbitals only, then for all core $p_{1/2}$ orbitals only,
and then for all core $p_{3/2}$ orbitals. As we would expect,
Uehling corrections to the core $s$ levels give most of the valence
shift, though consideration of the contributions from the $p$ orbitals 
is also important. We have found that the contribution from all $d$ 
orbitals of the core to the valence shifts for Cs gives a value that
is less than $10^{-4}$ times the size of the shifts from the core $s$
levels; that is, {\it $d$ level valence Uehling shifts are completely
  determined by the Uehling corrections to the $s$ and $p$ orbitals of
  the core}. 

Of the core $s$ levels, the uppermost core
$s$ orbital ($5s$ for Cs) gives the largest contribution, many times 
larger than the contribution from $1s$. 

This consideration is of significance for the problem of the 
radiative potentials for use in many-electron atoms, where
fitting factors are found by reproducing self-energy shifts in
hydrogen-like ions. To good accuracy, radiative potentials need only
be fitted to $s$ and $p$ self-energy shifts in hydrogen-like ions; it
is not necessary to fit to $d$ shifts. 
Also, it is more
important to optimize the fits for the higher, rather than the lower,
core $s$ levels. 

Due to the presence of $s$ electrons in the core, one may expect that QED radiative shifts
for $d$-levels for highly-charged ions are also determined largely
from core relaxation effects. This may change the magnitude and
sign of results of exact QED calculations for $d$-level shifts
performed in frozen atomic cores, e.g., in the work \cite{sapchengd}.

Our results for Na through to E119 are presented in
Table~\ref{tab:all-order}. First-order shifts, shifts with core
relaxation included, and shifts with core relaxation and electron-core 
correlations (to be discussed in the next section) are given. 
We see similar corrections for the other atoms that we saw for Cs.
The effect of relaxation on $s$ levels is significant but moderate,
about $(5-20)\%$. 
On the other hand, for all other levels, the magnitude and sign of the 
Uehling shift is determined by the core relaxation correction.
There is orders of magnitude enhancement of the $d$ level shifts for
all alkali atoms studied. For Cs, the $d$ levels experience a 
four-orders of magnitude enhancement. There is an almost 8 orders of magnitude
enhancement for the $3d_{5/2}$ level in Na. The relative size of this 
correction tends to decrease for atoms with higher $Z$, though remains 
very sizeable everywhere. For Rb, Cs, and Fr, the Uehling
shifts for $d$ levels are larger than those for $p$ levels. For Cs,
the $d$ level shifts are only ten times smaller than the
$s$ level shifts.

In Table~\ref{tab:all-order}, the results of Derevianko {\it et al.}
for $6s$ and $6p$ levels of Cs are presented in the final column; 
it is seen that our relativistic Hartree-Fock results agree very well.

The core relaxation corrections to $6s$ and $6p$ levels of Cs arising from
inclusion of the radiative potential (accounting for vacuum
polarization and self-energy) were calculated in Ref.~\cite{rad_pot}
and the importance of these effects were noted. The very large
corrections for $d$ levels were seen in our recent work on 
the spectra of the alkaline earths \cite{radium}. 
The relaxation effect has also been shown to be important in
the transition frequencies of several- and 
many-electron highly-charged ions \cite{blundell,CCJ}, particularly for
transitions involving $d$ levels \cite{4p4d}, 
although the relative size of the corrections seen there is 
significantly smaller than what we have observed in neutral atoms.

\section{Correlation corrections}
\label{sec:correlations}

\begin{table*}
\caption{First-order Uehling corrections to the binding energies
  for Cs, $\delta \epsilon^{(1)}_{\rm Br} =-\langle \varphi_{\rm Br} |V_{\rm
    Ueh}|\varphi_{\rm Br}\rangle$, where $\varphi_{\rm
    Br}$ is a solution of the Brueckner
  equation $(h_{\rm HF}+f_{\kappa}\Sigma_{\rm \kappa})\varphi_{\rm Br}
  =\epsilon\varphi_{\rm
  Br}$, and the correlation potential $\Sigma$ is the second order
$\Sigma ^{(2)}$ or all-orders $\Sigma ^{(\infty)}$. With no fitting,
$f_{\kappa}=1$ and $\epsilon=\epsilon_{\rm Br}$, while with fitting
$\epsilon=\epsilon_{\rm Exp}$. 
The numbers in square brackets $[\,]$ denote powers of 10. Units: a.u.}
\label{tab:sigma}
\begin{ruledtabular}
\begin{tabular}{lccccccc}
State &$\epsilon_{\rm
  Exp}$\tablenotemark[1]&\multicolumn{3}{c}{Second-order correlation
  potential $\Sigma^{(2)}$}&\multicolumn{3}{c}{All-orders correlation
  potential $\Sigma^{(\infty)}$}\\
 &&$\epsilon_{\rm Br}$&$\delta \epsilon^{(1)}_{\rm Br}$ &$\delta \epsilon^{(1)}_{\rm Br,fit}$ &
 $\epsilon_{\rm Br}$&$\delta \epsilon^{(1)}_{\rm Br}$ &$\delta \epsilon^{(1)}_{\rm Br,fit}$ \\
\hline
$6s_{1/2}$&-0.143098&-0.147671&-1.588[-5] &-1.470[-5]&-0.143262&-1.452[-5]&-1.448[-5]\\ 
$6p_{1/2}$ &-0.092166&-0.093578&-3.071[-7]&-2.856[-7]&-0.092436&-2.895[-7]&-2.855[-7]\\
$6p_{3/2}$ &-0.089642&-0.090849&-3.338[-8]&-3.124[-8]&-0.089848&-3.164[-8]&-3.129[-8]\\
$5d_{3/2}$&-0.077035&-0.080029&-4.970[-10]&-4.409[-10]&-0.078015&-4.605[-10]&-4.421[-10]\\
$5d_{5/2}$&-0.076590&-0.079296&-1.135[-10]&-1.015[-10]&-0.077501&-1.059[-10]&-1.018[-10]\\
\end{tabular}
\end{ruledtabular}
\tablenotetext{Data from NIST, Ref. \cite{NIST}.}
\end{table*}
 
In this section we consider the effect on the valence Uehling shift
due to account of higher orders in many-body perturbation theory. 
The many-body perturbation theory is expanded in the residual 
Coulomb interaction, 
$1/|{\bf r}_i-{\bf r}_j|+V_{\rm el}$, where $V_{\rm el}$ is the electronic potential used for the 
zeroth-order calculations.  
When calculations are performed in zeroth-order in the relativistic 
Hartree-Fock approximation, $V_{\rm HF}=V_{\rm el}$, as they have been
in this work, then the first-order many-body corrections cancel
exactly.

The lowest-order many-body corrections then arise in the 
second order in the Coulomb interaction. 
The corresponding diagrams, in the Goldstone and Feynman formalisms,
may be found, e.g., in Refs. \cite{review,dzuba}.
There is a method, developed by
Dzuba, Flambaum, and Sushkov \cite{DFS1989}, for taking into 
account the higher orders of many-body perturbation theory using the Feynman
diagram technique; the Coulomb lines are modified by including an infinite series of core
polarization loops and an infinite series of hole-particle interactions in each loop. 

A non-local, energy-dependent potential may be determined, with its 
averaged value corresponding to the 
correlation correction to the
energy. For example, for the second-order case $\delta \epsilon_i^{(2)}=\langle
\varphi_i|\Sigma^{(2)}({\bf r}_1,{\bf r}_2,\epsilon_i)|\varphi_i\rangle$ while
for the all-order case  $\delta \epsilon_i^{(\infty)}=\langle
\varphi_i|\Sigma^{(\infty)}({\bf r}_1,{\bf
  r}_2,\epsilon_i)|\varphi_i\rangle$. 
This potential, referred to as the {\it correlation potential}, may be 
added to the Hartree-Fock potential 
in the relativistic Hartree-Fock equations for the valence electrons. This
yields Brueckner orbitals $\varphi_{{\rm Br},i}$ and energies
$\epsilon_{{\rm Br},i}$. Such a procedure also takes 
into account the higher orders in $\Sigma$ in the Brueckner orbitals 
and energies.

In this work we calculate $\Sigma^{(2)}$ for all atoms considered and 
we calculate $\Sigma^{(\infty)}$ for Cs. 
We calculate $\Sigma^{(2)}$  using a B-spline basis set \cite{johnson}
obtained by diagonalizing the relativistic Hartree-Fock operator on a
set of 40 splines of order $k=9$ within a cavity of radius $40\,{\rm a.u.}$ 
The exchange part of $\Sigma^{(\infty)}$ 
is also considered in the second-order, with (multipolarity-dependent)
factors used to screen the Coulomb interaction. For the direct part of 
$\Sigma^{(\infty)}$, the Feynman diagram technique is used for
inclusion of the core polarization and hole-particle classes of
diagrams; see Ref.~\cite{dzuba} for further details about the method. 
The correlation potential method, using all-orders
$\Sigma^{(\infty)}$, has proven to be remarkably successful in
high-precision atomic structure calculations for heavy alkali atoms; 
notably, this method was used to obtain one of the most precise
results for parity violation in Cs \cite{DFS1989APV,DFG2002,APV2012}. 

Inclusion of the correlation potential modifies the orbitals of
the valence electrons at large distances, $r\gtrsim a_B$. These are the
distances that also determine the binding energies of the
electrons. Placing fitting factors before the correlation
potential $f_{\kappa}\Sigma_{\kappa}$ and fitting to the experimental 
binding energies may be considered a way of taking into account 
missed effects in the correlation potential. (Experience has shown 
that, indeed, fitting of the energies leads to improved wave
functions at all distances.) 

In Table \ref{tab:sigma} we present the Uehling shift for the valence 
orbitals of Cs with the correlation potential taken into account. We 
present results both for $\Sigma^{(2)}$ and $\Sigma^{(\infty)}$ to
see how sensitive the Uehling correction is to the correlation
corrections. The values presented in this table do not include the 
large effects of core relaxation. 

It is seen from Table \ref{tab:sigma} that the effect of the correlations is
large. They increase the $6s$ level shift by around $50\%$ and they 
exceed the relaxation correction for this level. The relative corrections for
the other waves is even larger, though for these waves the 
relaxation effect dominates.

\begin{table*}
\caption{The Uehling correction to binding energies. 
Zeroth-order relativistic Hartree-Fock binding 
energies are given in the third column. First order valence 
  corrections $\delta \epsilon^{(1)} =-\langle \varphi |V_{\rm
    Ueh}|\varphi\rangle$ and shifts including 
core relaxation $\delta \epsilon$ are given in the following columns. 
The values in the sixth column correspond to the addition of fitted
$\Sigma^{(2)}$ (for E119, fitting factors from Fr are used) to the
relativistic Hartree-Fock equations; the
shift is found from the difference in energies when Uehling is
included and excluded.
The numbers in square brackets $[\,]$ denote powers of
10. Units: a.u.}
\label{tab:all-order}
\begin{ruledtabular}
\begin{tabular}{llccccl}
Atom & State & $\epsilon_{\rm HF}$ & $\delta \epsilon^{(1)}$ & $\delta \epsilon$ & $\delta
\epsilon_{\rm Br, fit}$ &  Other \\
\hline
Na & $3s_{1/2}$ &-0.182033& -5.559[-7] &-5.910[-7] &-6.701[-7] &
-6.016[-7]\tablenotemark[1]\\ 
&$3p_{1/2}$ & -0.109490& -2.006[-10]& 3.067[-8] & 3.537[-8] & \\ 
&$3p_{3/2}$ &-0.109417&-4.101[-11]& 3.077[-8]& 3.548[-8]  & \\ 
&$3d_{3/2}$ &-0.055667&-6.538[-18] &-1.323[-10]& -1.412[-10]  &\\
&$3d_{5/2}$ &-0.055667 &-2.018[-18]& -1.347[-10]&-1.439[-10] &\\
K & $4s_{1/2}$ &-0.147491&-1.224[-6]&-1.333[-6] &-1.717[-6] &-1.371[-6]\tablenotemark[1]\\
&$4p_{1/2}$ &-0.095713&-1.879[-9]&7.262[-8]&9.376[-8] & \\
&$4p_{3/2}$ &-0.095498&-3.547[-10]&7.396[-8]&9.524[-8] & \\
&$3d_{3/2}$ &-0.058067&-1.046[-14]&1.395[-8]&4.172[-8] & \\
&$3d_{5/2}$ & -0.058080&-3.112[-15]&1.376[-8] &4.125[-8 ] & \\
Rb & $5s_{1/2}$ &-0.139291&-4.648[-6]&-5.114[-6]&-6.767[-6]&-5.288[-6]\tablenotemark[1] \\
&$5p_{1/2}$ & -0.090816&-3.234[-8]&1.776[-7]&2.365[-7]& \\
&$5p_{3/2}$ &-0.089986 &-4.881[-9]&2.108[-7]&2.774[-7]& \\
&$4d_{3/2}$ &-0.059687&-3.865[-12]&1.290[-7]&3.184[-7]& \\
&$4d_{5/2}$ &-0.059745&-1.048[-12]&1.218[-7]&3.001[-7]& \\
Cs & $6s_{1/2}$&-0.127368&-1.054[-5]&-1.160[-5]&-1.583[-5]&-1.206[-5]\tablenotemark[1], -1.054[-5]\tablenotemark[2], -1.159[-5]\tablenotemark[3]\\
&$6p_{1/2}$ & -0.085616 & -1.942[-7]&2.288[-7]&3.168[-7]&-1.942[-7]\tablenotemark[2], 2.284[-7]\tablenotemark[3] \\
&$6p_{3/2}$ &-0.083785 &-2.178[-8] & 4.518[-7] &6.097[-7]&-2.180[-8]\tablenotemark[2], 4.513[-7]\tablenotemark[3] \\
&$5d_{3/2}$ &-0.064420&-1.938[-10] &1.127[-6] &2.516[-6]& \\
&$5d_{5/2}$ &-0.064530&-4.612[-11]&1.007[-6] &2.216[-6]& \\
Fr & $7s_{1/2}$&-0.131076 & -4.997[-5]&-5.540[-5]&-7.227[-5]&-5.828[-5]\tablenotemark[1]\\
&$7p_{1/2}$ &-0.085911 &-2.569[-6]&-1.610[-6]&-2.414[-6]&  \\
&$7p_{3/2}$ &-0.080443&-1.356[-7]&1.834[-6]&2.408[-6]&  \\
&$6d_{3/2}$ &-0.062993&-3.579[-9]& 3.822[-6]&8.765[-6]&  \\
&$6d_{5/2}$ &-0.063444&-6.921[-10]&2.606[-6]&5.720[-6]&  \\
E119 & $8s_{1/2}$ &-0.152842 &-3.427[-4]& -4.046[-4]&-4.484[-4]& -4.436[-4]\tablenotemark[1]\\
&$8p_{1/2}$ & -0.091697 &-3.921[-5]&-4.701[-5]&-7.643[-5]& \\
&$8p_{3/2}$ &-0.075972&-5.420[-7]&1.120[-5]&1.430[-5]&  \\
&$7d_{3/2}$ &-0.061414&-2.517[-8]&8.715[-6]&2.594[-5]&  \\
&$7d_{5/2}$ &-0.063000 &-4.420[-9] &-1.374[-6]&9.042[-7]& \\
\end{tabular}
\end{ruledtabular}
\tablenotetext[1]{Relativistic HF, perturbative treatment of vacuum polarization, Ref. \cite{thierfelder}.}
\tablenotetext[2]{First-order relativistic HF results, $\delta
  \epsilon^{(1)}$, Ref. \cite{derev_relax}.}
\tablenotetext[3]{Relativistic HF with relaxation included, $\delta
  \epsilon$, Ref. \cite{derev_relax}.}
\end{table*}

The effect of the correlation potential is to pull the wave
functions closer to the nucleus. This leads to larger Uehling 
shifts for all waves and is more important for the orbitals with 
$l>0$.

The difference between the results using $\Sigma^{(2)}$ or
$\Sigma^{(\infty)}$ is significant, affecting the second digit in the
results. However, the fitted results 
shown in columns five and eight of 
Table \ref{tab:sigma}, agree very well, to about 2\% for $6s$ and significantly better
for the other waves. That is, {\it as long as the correlation potential is 
fitted to reproduce the experimental energies, the correlation
corrections to the Uehling shifts are insensitive to the details of $\Sigma$}.

Therefore, we have performed calculations for Na through to E119 with
the second-order correlation potential $\Sigma^{(2)}$ fitted to
reproduce experimental energies (for E119, we use fitting factors
from Fr). These results, which also include 
the important core-relaxation effects, are presented in the sixth
column of Table \ref{tab:all-order}. 

In the final column of Table \ref{tab:all-order} we present, along
with the results of Ref.~\cite{derev_relax}, the vacuum polarization 
calculations performed by Thierfelder and Schwerdtfeger
\cite{thierfelder} in the relativistic Hartree-Fock approach. 
Their results are in reasonable agreement with our relaxed core results $\delta \epsilon$.  
We are not aware of other many-body vacuum polarization results for the alkali atoms.

\begin{figure}[htbp]
\begin{center}
\includegraphics[width=\columnwidth]{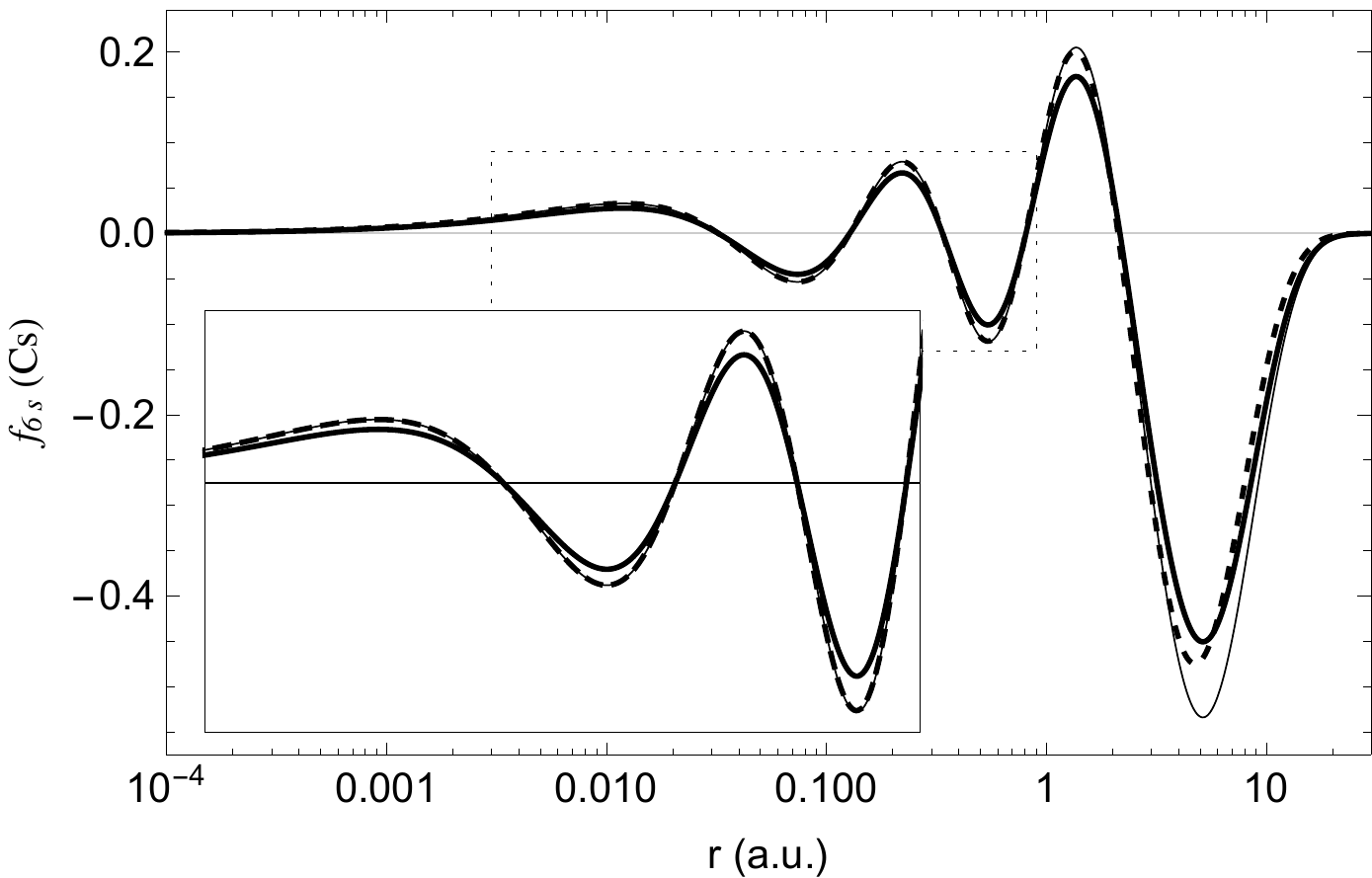}
\caption{Upper radial component of the Cs 6s orbitals. Thick solid
 line corresponds to the relativistic HF orbital, dashed line is the Brueckner 
orbital, and the thin solid line is the rescaled relativistic HF orbital, which
is seen to coincide with the Brueckner orbital for $r \lesssim a_B$
(see inset).}
\label{fig:Cs6sOrbitals}
\end{center}
\end{figure}

It is worth emphasizing the effect of the correlation potential on the
valence electron wave functions. The wave functions are modified by the correlation
potential in the region $r\gtrsim a_B$; see
Fig.~\ref{fig:Cs6sOrbitals}, where we illustrate the relativistic Hartree-Fock and
Brueckner upper radial component for the $6s$ orbital of Cs. 
For $r<a_B$, the Brueckner
wave functions are proportional to the Hartree-Fock wave functions,
$\varphi_{\rm Br}(r<a_B)=c\varphi(r<a_B)$. The effect that the
correlation corrections have in this region is simply in the
normalization of the wave functions. Obviously, since the
Uehling potential acts at a distance close to the nuclear radius,
$\langle \varphi_{\rm Br}|V_{\rm Ueh}|\varphi_{\rm Br}\rangle =c^2\langle
\varphi|V_{\rm Ueh}|\varphi\rangle$. If the range of the correction to
the Hartree-Fock potential $\delta V_{\rm HF}^{\rm Ueh}$ were within $a_B$,
then $\langle \varphi_{\rm Br}|V_{\rm Ueh}+\delta V_{\rm HF}^{\rm Ueh}|\varphi_{\rm Br}\rangle =c^2\langle
\varphi|V_{\rm Ueh}+\delta V_{\rm HF}^{\rm Ueh}|\varphi\rangle$ also,
i.e., $\delta \epsilon _{\rm Br}=c^2 \delta \epsilon= (\delta
\epsilon^{(1)}_{\rm Br}/\delta \epsilon^{(1)})\delta \epsilon$. We do notice 
some deviation from this relation, which is a result of the long range of
the relaxation correction to the Hartree-Fock potential; this range indeed
extends beyond $r=a_B$, as we saw in the previous section. Overall,
though, the relation above holds fairly well.

From consideration of the level of agreement between the fitted 
Brueckner results of Cs from Table \ref{tab:sigma}, we estimate the 
uncertainty of our final results for the alkali atoms Na to E119 
(column six of Table \ref{tab:all-order}) to be a few percent.

While the aim of this work is to study the effect of many-body
corrections on the QED radiative shifts, applied to the Uehling 
potential specifically, we note that there are other 
contributions to the vacuum polarization shift at the one-loop
level that we have not included, 
for instance the higher-order in $Z\alpha$ corrections 
and remaining electron screening effects corresponding to a vacuum polarization
loop in the photon exchange between electrons.
A more precise study would include these contributions, 
though we expect them to contribute at a level
that is smaller than the estimated uncertainty in this work.

\section{Conclusion}

We have studied the mechanisms of core relaxation and valence-core correlations
on the Uehling shift for the valence levels of the alkali
atoms Na to E119. We observe sizeable corrections to the $s$-level shifts, while
for the other orbitals $l>0$ the relative size of the corrections is
larger. For the $p_{1/2}$ orbitals we have seen for several atoms that
the relaxation and first-order Uehling shifts are roughly the same
size and of opposite sign. For $p_{3/2}$ and for the $d$ levels, the
effect of core relaxation is enormous, with corrections being 
orders of magnitude larger than the first-order result and usually of
opposite sign. Account of the valence-core correlations is also important for all
waves. For Rb, Cs, and Fr, the $d$-level shifts become
comparable to the size of the $s$-level shifts (one order of magnitude
smaller).

The atomic theory uncertainty for calculations of transition
frequencies in heavy neutral alkali atoms is at the level of 0.1\%
\cite{dzuba}, limited by the incomplete treatment of electron-electron correlations.  
This is roughly the same level where the radiative corrections enter.
(The full radiative shift is comprised, at one-loop level, of the vacuum
polarization and self-energy shifts, with the self-energy shift typically an order of
magnitude larger than the vacuum polarization shift and of opposite sign.)
If the remaining uncertainty in the electronic theory can be controlled, 
the huge many-body corrections for the radiative shifts for 
$d$ levels could make high-precision studies of transition frequencies involving these levels 
a particularly sensitive test of combined QED and many-body effects.
 
The results of this work are relevant for studies of radiative
potentials -- potentials that mimic self-energy QED radiative
corrections -- for use in heavy atoms and ions. 
The many-body enhancement mechanisms that operate for the Uehling 
potential apply also to the self-energy due to the short range of the interaction.  

\section*{Acknowledgments}

We are grateful to A. Yelkhovsky for useful discussions. The all-order 
correlation potential $\Sigma^{(\infty)}$ was calculated using the 
Dzuba-Flambaum-Sushkov codes maintained by V. Dzuba. 
This work was supported in part by the Australian Research Council, 
grant DE120100399.

\end{document}